\documentclass{aastex}
\usepackage{spr-astr-addons}
\usepackage{url}

\RequirePackage{color}
\newcommand{\photu}{photons cm$^{-2}$ s$^{-1}$ sr$^{-1}$ \AA$^{-1}$}
\newcommand{\GALEX}{\textit{GALEX}}

\begin{document}

\title{The Diffuse Ultraviolet Foreground}
\author{Jayant Murthy\altaffilmark{1}}
\affil{Indian Institute of Astrophysics, Bangalore 560034, India}
\altaffiltext{1}{jmurthy@yahoo.com}

\begin{abstract}

Ultraviolet observations from low Earth orbit (LEO) have to deal with a foreground comprised of airglow and zodiacal light which depend on the look direction and on the date and time of the observation. We have used all-sky observations from the GALEX spacecraft to find that the airglow may be divided into a baseline dependent on the sun angle and a component dependent only on the time from local midnight. The zodiacal light is observable only in the near ultraviolet band (2321 \AA) of GALEX and is proportional to the zodiacal light in the visible but with a color of 0.65 indicating that the dust grains are less reflective in the UV.

\end{abstract}


\section{Introduction}

Measurements of the diffuse ultraviolet (UV) radiation field have to contend with a number of contaminating sources including atmospheric emission lines and the zodiacal light \citep{murthy2009}. These foreground sources are particularly important at high galactic latitudes where the Galactic contribution to the radiation field is relatively small and at longer wavelengths where the zodiacal light, which follows the solar spectrum, becomes increasingly important. It has been difficult to disentangle these components, largely because of a lack of relevant observations. Ideally, these would be spectroscopic observations with moderate resolutions over a large part of the sky with different sun angles. However, what we have is thousands of observations from the Galaxy Evolution Explorer (\GALEX) in two bands (FUV: 1531 \AA\ and NUV: 2321 \AA) with observations far from the Sun to minimize foreground emission.

Despite these drawbacks, we have used the \GALEX\ data  to derive empirical formulae for the foreground sources. Although our main interest is in better understanding the galactic and extragalactic diffuse radiation, we hope that our results will also prove useful in studies of the  Earth's atmosphere and of the zodiacal light. They will certainly prove useful in mission planning for other space-borne instruments such as the Ultraviolet Imaging Telescope \citep{kumar2012} which will observe the sky with large field of view instruments where diffuse radiation limits the observable sky.

\section{Observations \& Data}

\subsection{Observations}

The \GALEX\ spacecraft was launched in 2003 and has since observed about 75\% of the sky in two spectral bands (FUV: 1531 \AA\ and NUV: 2321 \AA) with a spatial resolution of about 5\arcsec\ over a field of view of 0.6\degr. The primary mission was described by \citet{martin2005} and the software and data products by \citet{morrissey2007}. Most of the \GALEX\ observations were short exposures of about 100 seconds in length (All-Sky Imaging Survey: AIS) but there were a number of longer observations of 10,000 seconds or more, either to fulfill specific mission objectives or taken as part of the Guest Investigator (GI) program. A single exposure was limited by the duration of the orbital night (about 1000 seconds) and longer observations were broken up into a series of exposures spread over a time period ranging from days to years. \GALEX\ observations were subject to severe selection effects due to brightness related constraints from the diffuse radiation integrated over the large field of view. Thus the Galactic plane and other high intensity regions such as Orion or the Magellanic Clouds could not be observed in the original mission. Recent observations have covered many of these but they will not help in refining the foreground because of the brightness of the astrophysical emission. Observations were only taken at orbital night between 20:00 and 04:00 (local time) and only in directions more than 90\degr\ from the Sun to minimize airglow and zodiacal light. Hence, we only sampled a limited part of the total phase space of observational parameters.

\begin{figure}[t]
\includegraphics[width=\columnwidth]{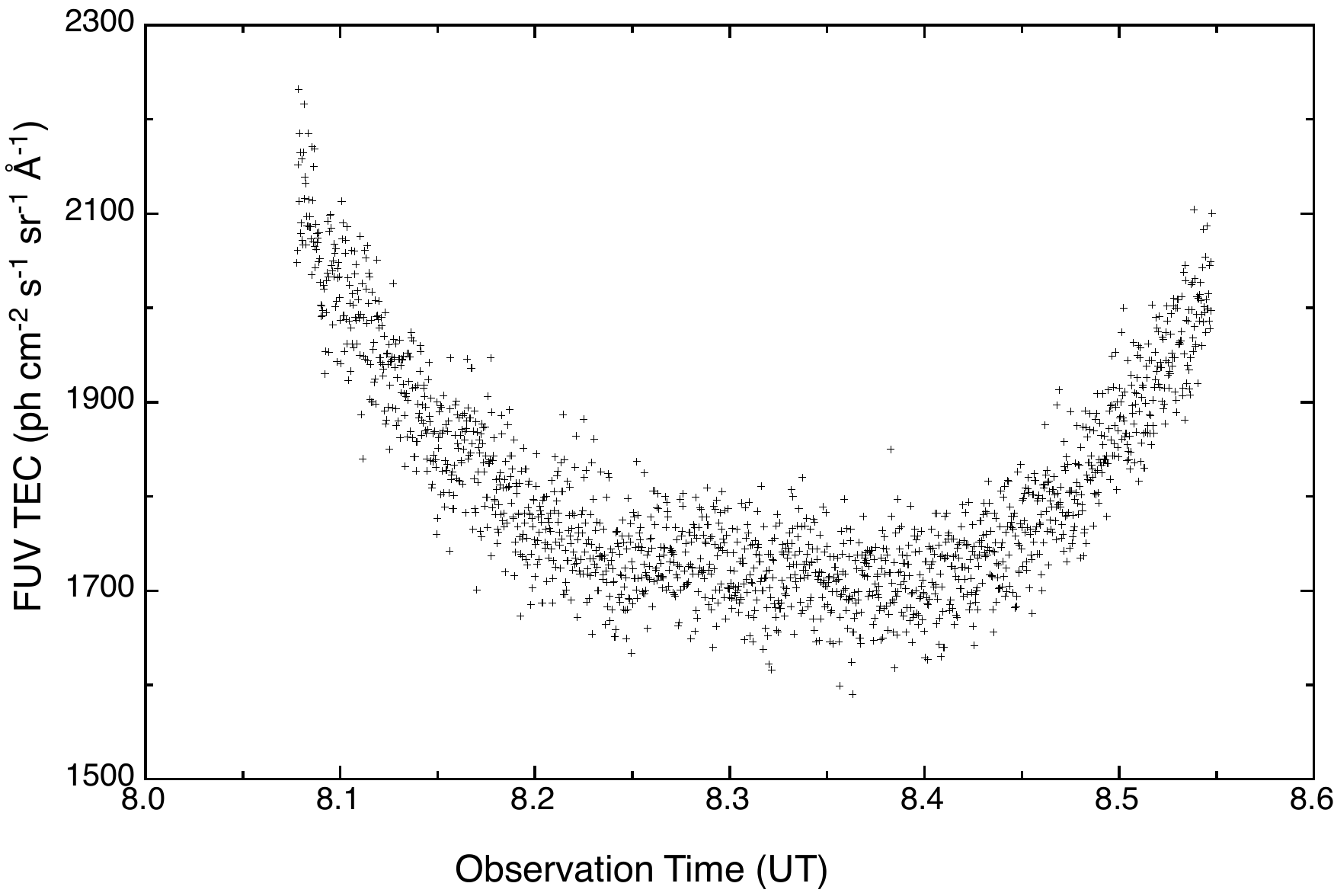}
\caption{\label{fig:tec_counter}The Total Event Counter (TEC) as a function of UT. The minimum emission was at local (spacecraft) midnight.}
\end{figure}

The diffuse background is the sum of the galactic background, which depends only on the look direction, and the foreground emission --- airglow and zodiacal light --- which depends on the time and date of the observation. The standard data products include a single image of the targeted field for each of the two bands (the FUV detector failed in May, 2009 following which observations were made only with the NUV detector) and a merged catalog of point sources from both bands. Because we are trying to derive the foreground emission which is a function of time and date of the observation, we used the spacecraft housekeeping files (scst files) which included the total count rate (TEC: Total Event Count) in each of the two detectors. The TEC was tabulated every second and was tied to the UT (universal time) of the observation. A typical TEC is plotted with respect to UT in Figure \ref{fig:tec_counter}. 

%Figure tec_counter

As implied by its name, the TEC includes all emission in the field of view including starlight, diffuse background, airglow, and zodiacal light and each element had to be estimated separately. However, only the airglow would be expected to vary with the time of day and, to anticipate our results, specifically with the time from local midnight. Unfortunately, this was not readily available from the \GALEX\ data products and we obtained the TLEs (two-line elements) from Space-Track.org (\verb+https://www.space-track.org/+) and then used STK (\verb+http://www.agi.com/default.aspx+) to calculate the latitude and longitude of the spacecraft ground track at a given UT. From this we calculated the local spacecraft time and cross-indexed with the TEC to obtain the total emission as a function of time. As a result of this rather painful experience, we would recommend that future missions include the local spacecraft coordinates as part of their standard data products.

There were about 34,000 observations in each of the FUV and NUV bands in the \GALEX\ GR6 data release with close to 76,000 scst files and 14,000,000 individual TEC measurements in each band. Note that the longer exposures were divided into multiple exposures and hence multiple scst files.

\subsection{Airglow}

Naturally enough, most observations of the airglow from space have been downward looking and have found a number of different atomic and molecular lines (reviewed by \citet{meier1991}). There are many fewer observations looking up from low-Earth orbit (LEO). The only emission lines observable in the night spectrum are the geocoronal O I lines at 1304 \AA\ (0.013 kR) and 1356 \AA\ (0.001 kR) in the FUV band and 2471 \AA\ ($< 0.001$ kR) in the NUV band \citep{morrison1992,  feldman1992, boffi2007}. The \GALEX\ FUV band rejected the 1304 \AA\ line but with a 10\% leak \citep{morrissey2007} and these values corresponded to expected levels of about 200 \photu\ in the FUV band and 100 \photu\ in the NUV band.

As mentioned above, we have adopted an empirical approach in studying the foreground emission. We noted that each observation could be separated into two parts; a minimum value at orbital midnight and a time-variable part which increased smoothly on either side of orbital midnight (Figure \ref{fig:tec_counter}). The only possible source for a signal that varies with the local time is airglow and we extracted this component of the foreground emission by subtracting a baseline calculated from the average of the points within 15 minutes of local midnight, assuming that the observation included this time span. This baseline is comprised of all other sources in the field, including any residual airglow emission at midnight.

%Figure ag_density
\begin{figure*}[t]
\includegraphics[width=\columnwidth]{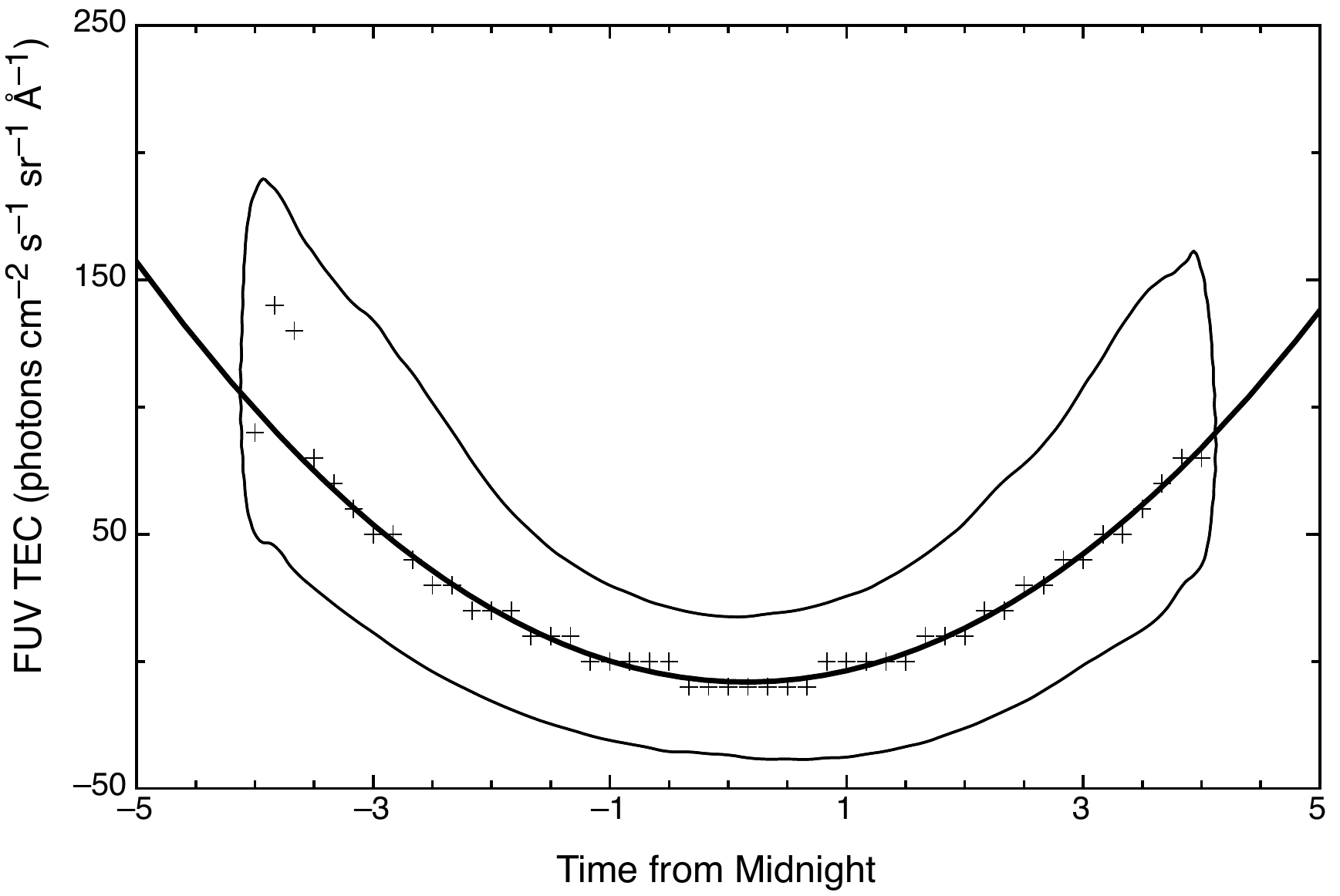}
\includegraphics[width=\columnwidth]{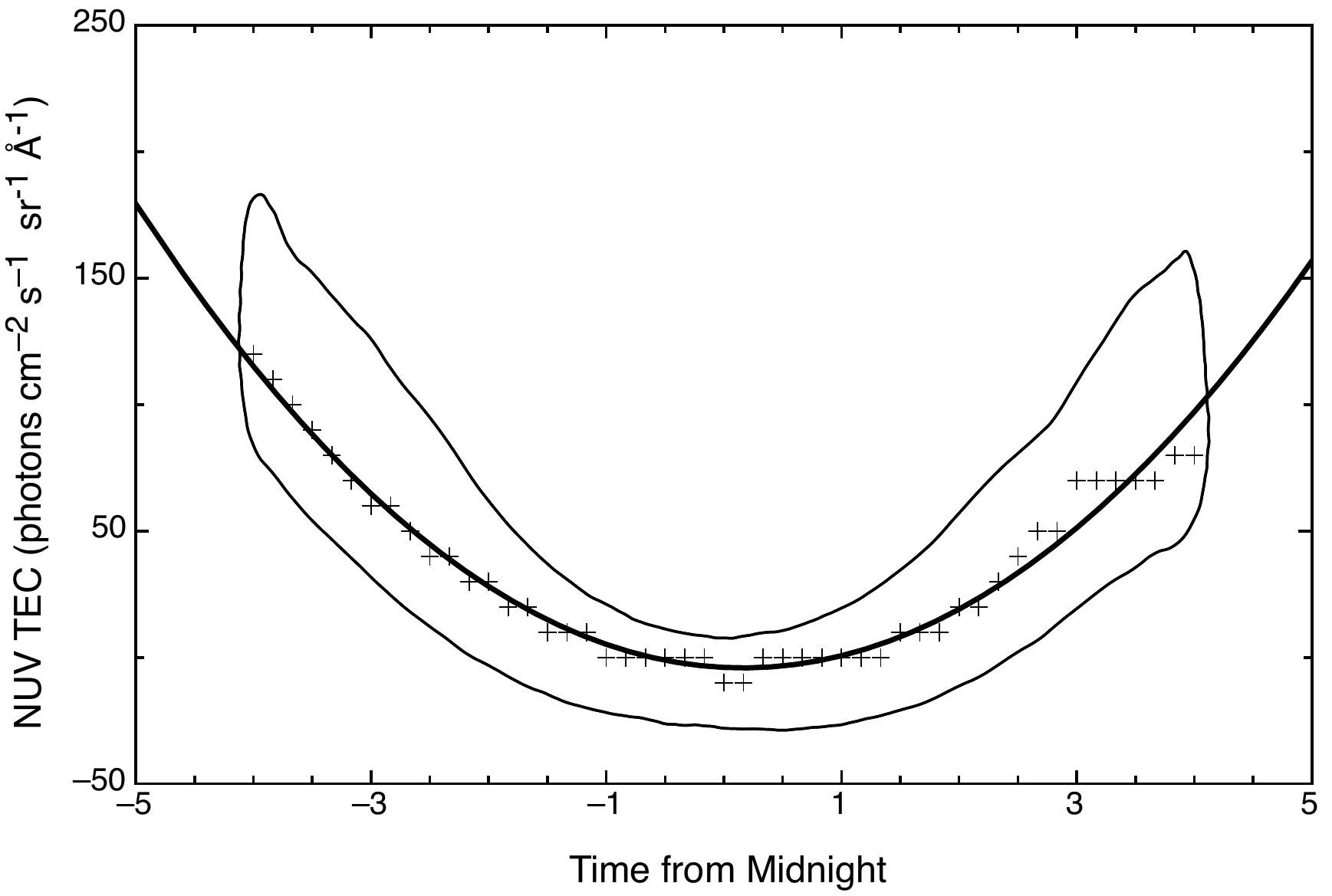}
\caption{Distribution of FUV (a) and NUV (b) baseline-subtracted airglow. Plus signs (+) show the peak of the distribution and the contour lines represent the 1$\sigma$ limits on the level of the airglow. The dark line is the parametrized fit to the peak airglow.}
\label{fig:ag_density}
\end{figure*}

There were approximately 5.8 million independent points in the FUV channel and 6.6 million in the NUV channel with an overlap of about 5 million points. We gridded the baseline-subtracted data to form a density plot for each band and these are shown in Figure \ref{fig:ag_density}. These plots were created by gridding the data into bins of 10 minutes in time and 10 \photu\ in flux. The plus signs show the peak airglow at a given time and the 1$\sigma$ contour is shown as the closed line; that is, 68\% of the data points fall within the two contour lines between the observation limits of 20:00 to 04:00. The dark lines are the best fit quadratics to the airglow emission in each band with the equations: $$FUV = -7.97 - 1.91 t + 6.22 t^2	(1)$$
$$NUV = -3.94 - 2.25 t + 6.88 t^2	(2)$$
where FUV and NUV are the total flux in the respective bands in units of \photu\ and t is the time from midnight in hours. If the observation stretches the entire 8 hours, the maximum for a \GALEX\ observation, the airglow contribution would be 25 and 33 \photu\ in the FUV and NUV bands, respectively, with an estimated error of about 20 \photu\, primarily due to the difficulty of measuring the baseline. Note that the time measured is time on the ground track where 8 hours corresponds to about 20 minutes of actual spacecraft time. There is a strong correlation between the FUV and NUV data (r = 0.745) with NUV = 0.7 FUV, consistent with the origin of both in geocoronal O I lines.

\subsection{Zodiacal Light}

The zodiacal light is due to sunlight scattered by interplanetary dust in the UV and visible and thermal emission from the dust in the infrared, with a black body temperature of about 60 K. It has been observed extensively in the visible from ground-based observations (reviewed by \citet{leinert1998}) and in the infrared \citep{kelsall1998} from the \textit{Infrared Astronomy Satellite} (\textit{IRAS}) mission. It is generally assumed that the spectrum of the zodiacal light follows the Solar spectrum; that is, the color of the zodiacal light is unity. However, there have been very few observations in the ultraviolet with the most robust being those of \citet{murthy1989} who claimed that the color of the zodiacal light increased with the ecliptic angle. In any case, the zodiacal light will only contribute in the NUV band.

\begin{figure}[t]
\includegraphics[width=\columnwidth]{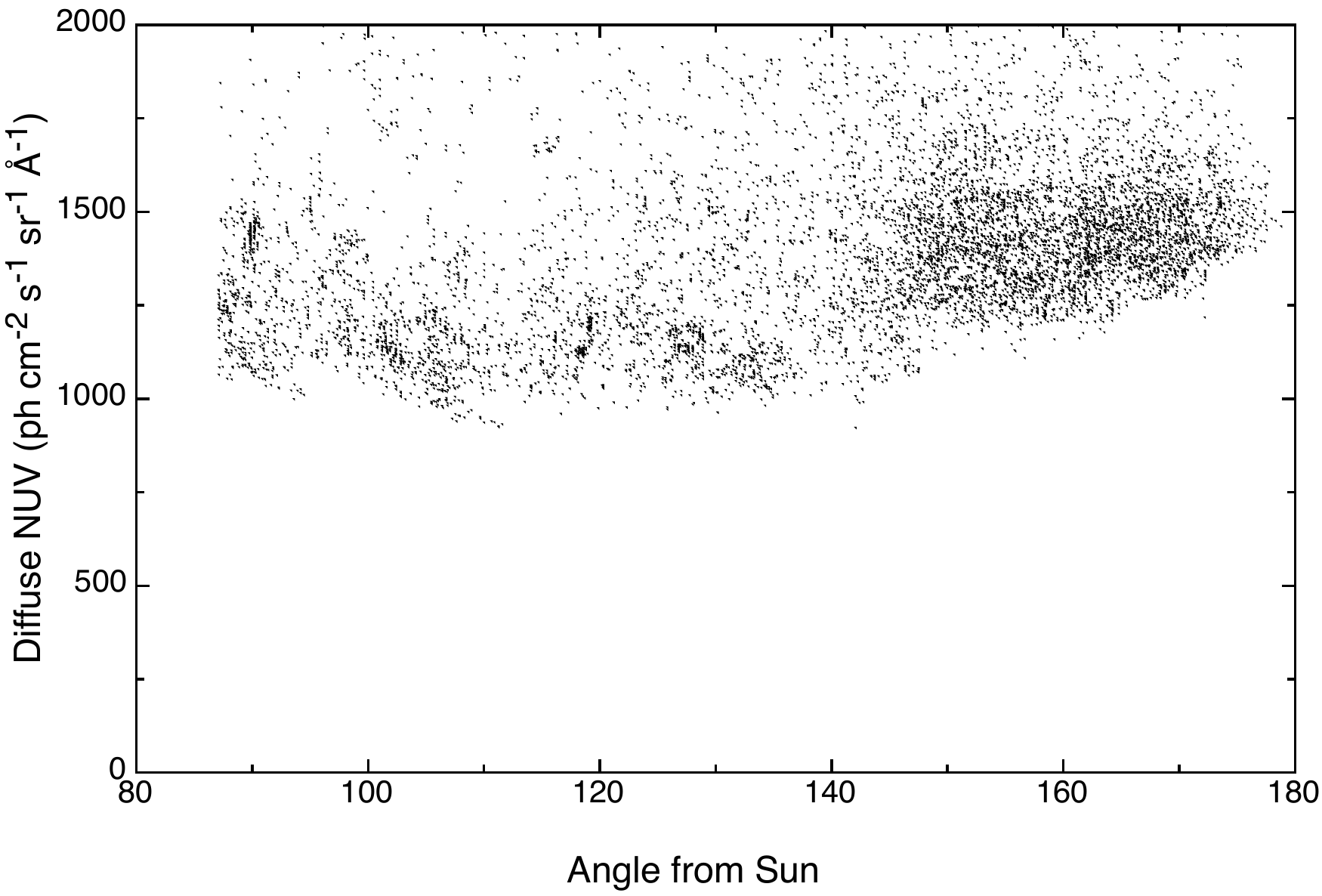}
\caption{Distribution of NUV TEC values at local midnight with the starlight subtracted. Airglow is responsible for the rise in the values in the left and zodiacal light for the rise in the right.}
\label{fig:nuv_baseline}
\end{figure}

We further restricted our dataset to those observations which included both FUV and NUV data and where the observations extended at least 15 minutes on either side of orbital midnight such that a baseline could be robustly defined. There were 9313 such observations and we extracted the baseline value for each. This baseline value at orbital midnight includes, as mentioned above, starlight, the galactic background, airglow and, in the NUV band only, the zodiacal light. The \GALEX\ pipeline produces a merged catalog for each observation in which the fluxes of the point sources in each band are tabulated from which we could estimate and subtract the contribution due to starlight in each band. These data --- the baseline value at midnight from which the starlight has been subtracted --- form the basis of our further analysis and the NUV data (including residual airglow; diffuse astrophysical background; and zodiacal light) are plotted as a function of sun angle in Figure \ref{fig:nuv_baseline}. There is, of course, considerable scatter reflecting the range in galactic latitude spanned by the observations but a lower envelope to the values can be cleanly drawn with the rise on the right due to locations at low ecliptic latitudes.

%Figure tec_counter

\begin{figure}[t]
\includegraphics[width=\columnwidth]{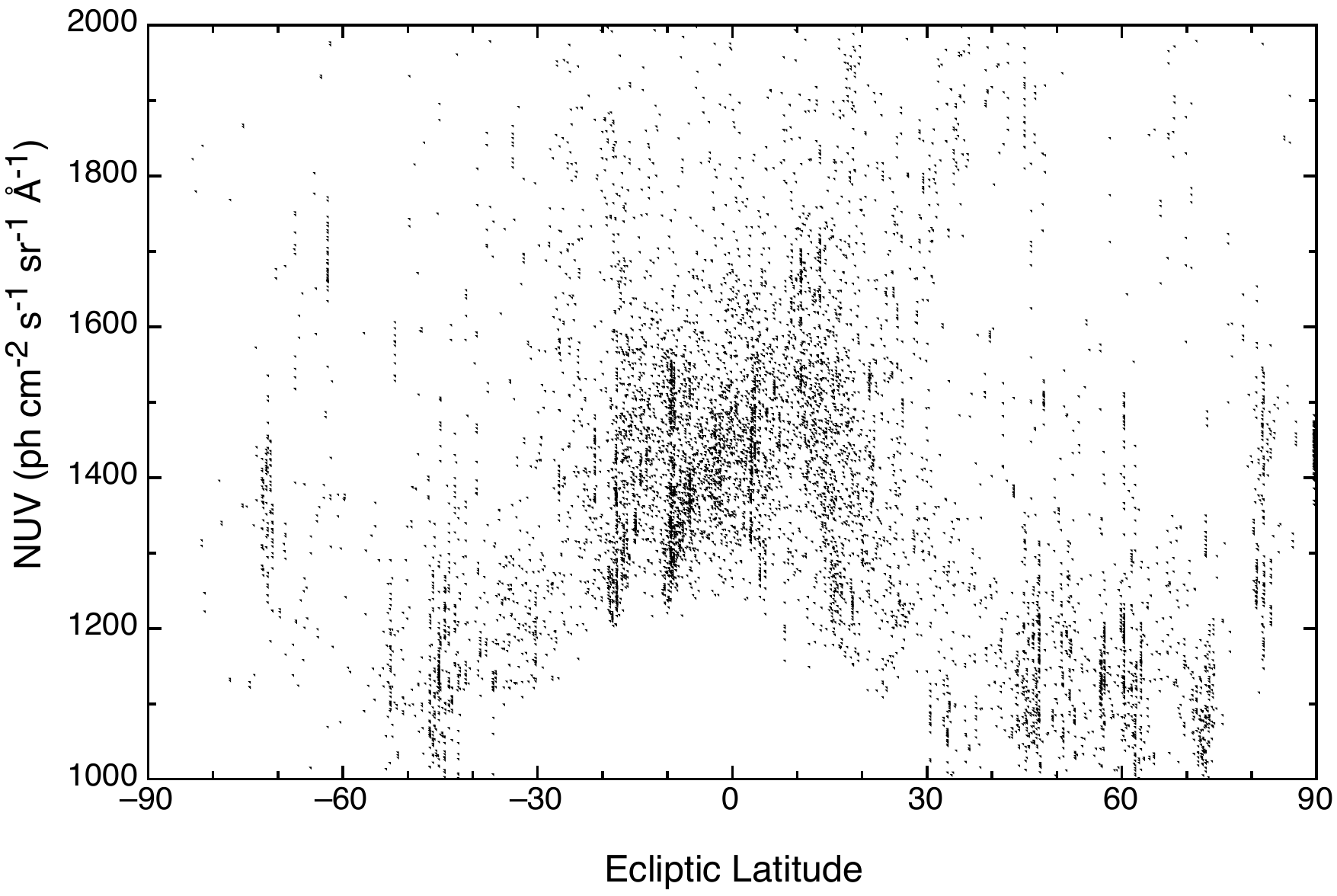}
\caption{\label{fig:zod_with_ecliptic}NUV TEC as a function of ecliptic latitude.}
\end{figure}

\begin{figure}[t]
\includegraphics[width=\columnwidth]{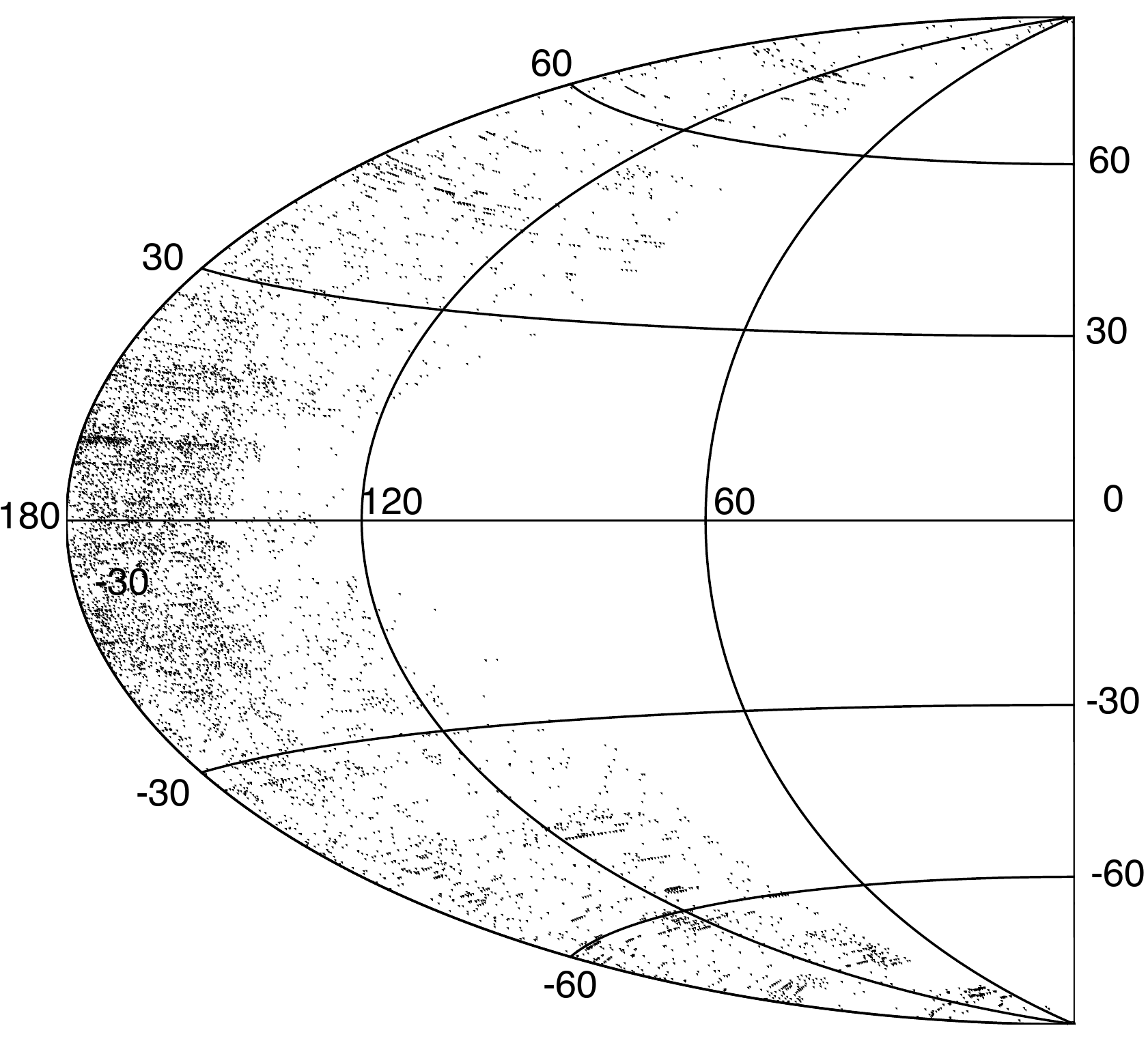}
\caption{\label{fig:obs_with_earth}Observations as a function of helioecliptic longitude and latitude.}
\end{figure}

\begin{figure}[t]
\includegraphics[width=\columnwidth]{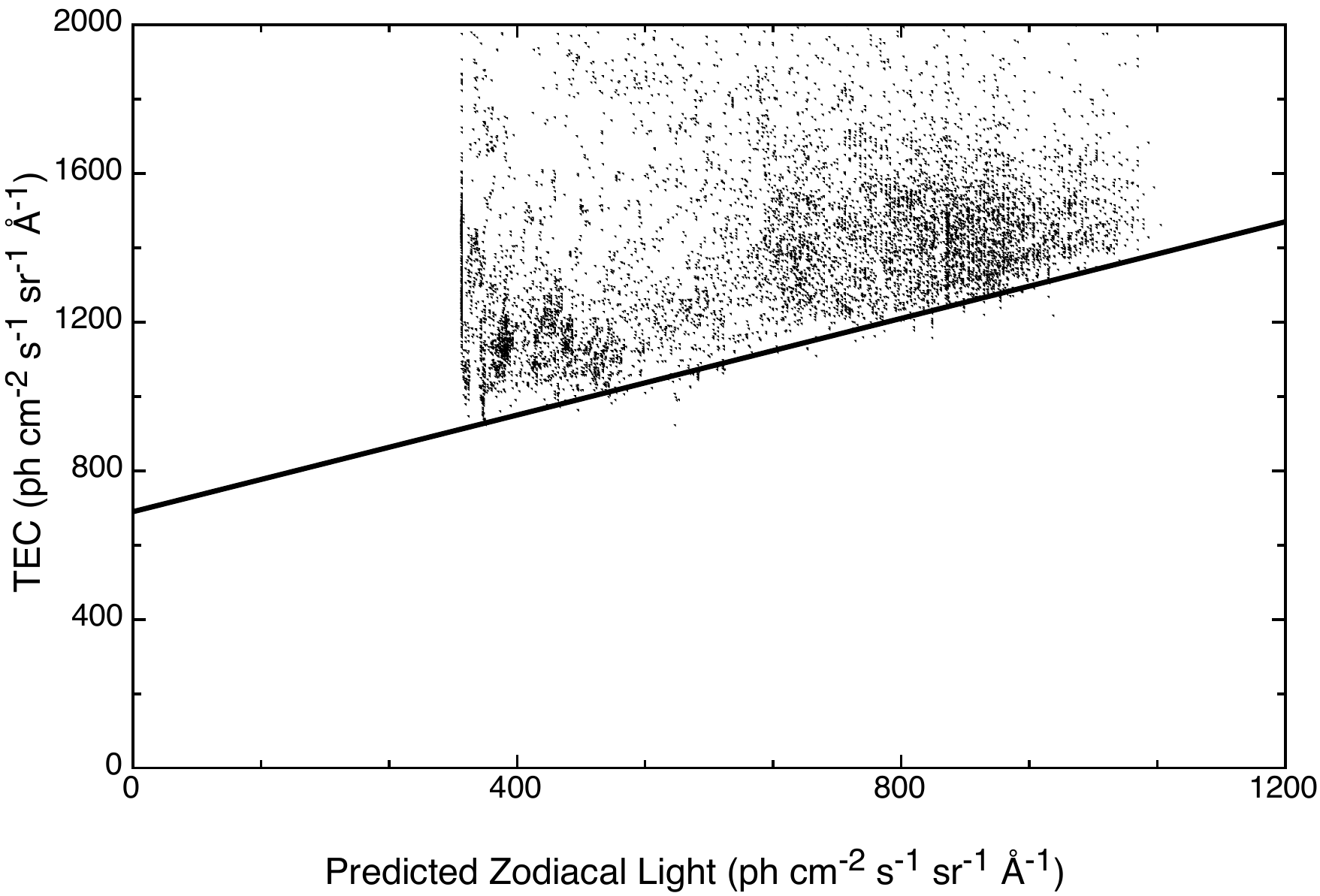}
\caption{\label{fig:zl_obs_mod}Observed NUV TEC as a function of predicted zodiacal light.}
\end{figure}

Zodiacal light does contribute significantly to the NUV channel  and is strongly dependent on the ecliptic latitude (Figure \ref{fig:zod_with_ecliptic}) and the helioecliptic longitude (ecliptic longitude - longitude of the Sun). Because of operational constraints on \GALEX\ observations, all the data were taken at large helioecliptic longitudes (Figure \ref{fig:obs_with_earth}) and, in fact, there was little variation with longitude. \citet{leinert1998} have tabulated observations of the zodiacal light in the optical and, as a first estimate, we have assumed that the zodiacal light in the UV follows the optical distribution with a spectral correction given by the solar spectrum (eg. \citet{colina1996}). The predicted zodiacal light tracks the observed TEC well (Figure \ref{fig:zl_obs_mod}) with the formula: TEC = 690 + 0.65 ZL, where ZL is the zodiacal light from the visible observations by \citet{leinert1998}. The baseline is the zero level representing the residual airglow and the diffuse galactic light while the slope represents the color of the zodiacal light with respect to the visible --- the scattered light in the UV is 65\% that in the optical. Unlike \citet{murthy1989}, we find no dependence of the color on the ecliptic latitude. Although a detailed study of the zodiacal dust is beyond the scope of this work, this suggests that the albedo (reflectivity) of the interplanetary dust grains in the UV is 65\% of the albedo in the visible, in rough agreement with results for interstellar dust \citep{draine2003}.

%Figure TEC with ecliptic latitude

%Figure TEC with ecliptic latitude

%Figure TEC with ecliptic latitude

\subsection{Airglow Redux}

\begin{figure*}[t]
\includegraphics[width=\columnwidth]{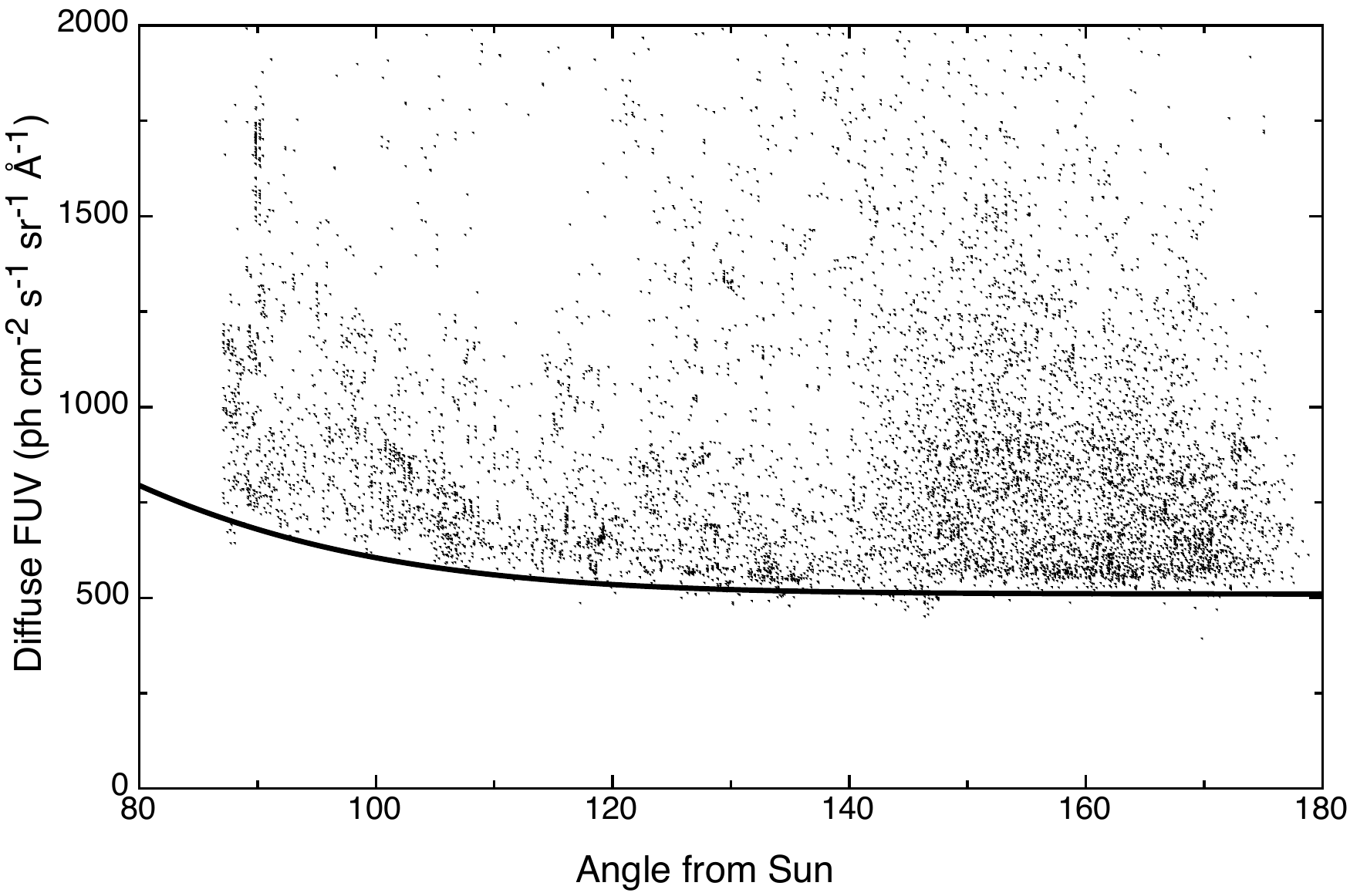}
\includegraphics[width=\columnwidth]{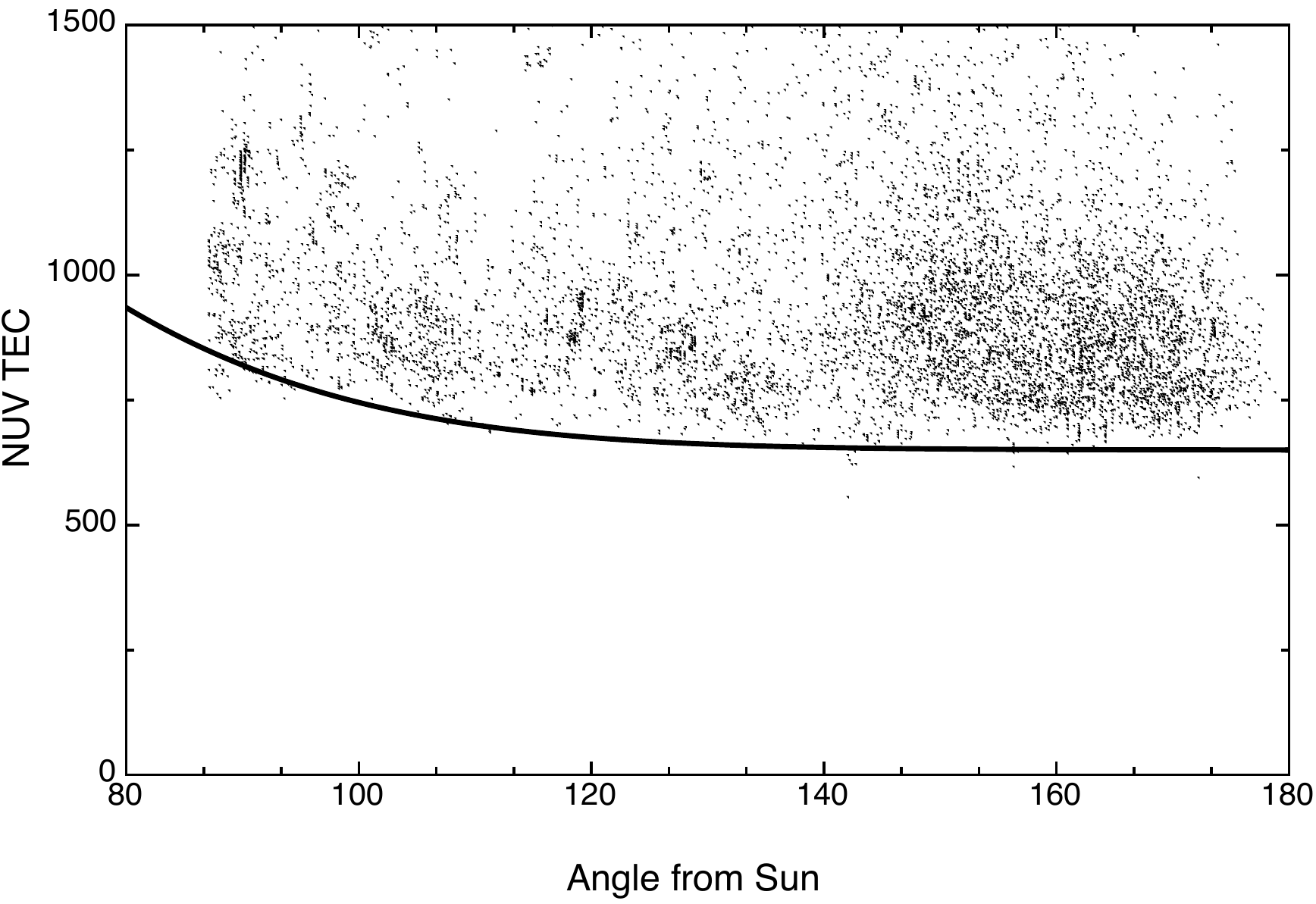}
\caption{Distribution of FUV (a) and NUV (b) TEC values at local midnight. The starlight has been subtracted in each case and, further, the zodiacal light was subtracted in the NUV case. The dark line is the parametrized lower envelope to the TEC.}
\label{fig:ag_baseline}
\end{figure*}

We have subtracted starlight from both bands and zodiacal light from the NUV band and plotted the dependence of these on the Sun angle in Figure \ref{fig:ag_baseline}.  Note that the rise in data values on the right side of Figure \ref{fig:nuv_baseline} is no longer seen, indicating that it is, indeed, due to zodiacal light. We have empirically defined a lower envelope to the FUV and NUV TEC by the equations $$FUV = 2000 e^{-SA^2} + 520\ (3)$$  $$NUV = 2000e^{-SA^2} + 650\ (4)$$ where SA is the angle from the Sun in radians and FUV and NUV are the respective TEC values in units of \photu. The variable component must be due to the airglow in the FUV, where zodiacal light will not contribute, and hence is likely to be from the airglow in the NUV also, where it  follows the same dependence on Sun angle. However, there is no way to independently separate the astrophysical component from the baseline airglow. If we take recourse to the canonical value of 300 \photu\ for the diffuse background \citep{henry2010}, we find that the FUV airglow is 220 \photu\ at midnight and the NUV airglow is 350 \photu.

\section{Conclusions}

We have empirically derived predictions for the foreground emission in the two \GALEX\ bands. There are two components to the airglow: one which is dependent on the local time and is symmetrical around local midnight; and the other which is dependent on the angle between the target and the Sun. We find that the zodiacal light is proportional to that in the visible with a color of 0.65. We are left with an ambiguity in the baseline at the 100 - 200 \photu\ level which will require spectroscopic information to disentangle.

\bibliographystyle{spr-mp-nameyear-cnd}
%\bibliography{myref}
%\bibliography{murthy_ms}

\end{document}